\documentclass[preprint,showpacs,titlepage,aps,prd,
tightenlines,byrevtex,nofootinbib]{revtex4}

\usepackage{graphicx,amssymb}

\def\GWth{\text{GW}_{\text{th}}}
\def\eV{{\text{eV}}}
\def\MeV{{\text{MeV}}}

\def\best{{\text{best}}}
\def\sys{{\text{sys}}}
\def\J-PARC{{\text{J-PARC}}}
\def\react{{\text{react}}}
\def\rCP{{\text{CP}}}
\def\CP1{{\text{CP1}}}
\def\tot{{\text{tot}}}

\def\sigDB{\sigma_{\text{DB}}}
\def\sigDb{\sigma_{\text{Db}}}
\def\sigdB{\sigma_{\text{dB}}}
\def\sigdb{\sigma_{\text{db}}}
\def\sigD{\sigma_{\text{D}}}
\def\sigd{\sigma_{\text{d}}}
\def\sigB{\sigma_{\text{B}}}
\def\sigb{\sigma_{\text{b}}}
\def\sigBG{\sigma_{\text{BG}}}
\def\sigS{\sigma_{\text{S}}}
\def\BG{{\text{BG}}}

\begin{document}

\preprint{hep-ph/0309323}

\title{Exploring Leptonic CP Violation by Reactor and 
Neutrino Superbeam Experiments
}

\author{Hisakazu Minakata}
\email{E-mail: minakata@phys.metro-u.ac.jp}
\author{Hiroaki Sugiyama}
\email{E-mail: hiroaki@phys.metro-u.ac.jp}
\affiliation{Department of Physics, Tokyo Metropolitan University,
Hachioji, Tokyo 192-0397, Japan}


\vglue 1.4cm

\begin{abstract}
We point out the possibility that reactor measurement of 
$\theta_{13}$, when combined with high-statistics $\nu_e$ 
appearance accelerator experiments, can detect leptonic 
CP violation. 
Our proposal is based on a careful statistical analysis under 
reasonable assumptions on systematic errors, assuming 2 years 
running of the neutrino mode J-PARC $\rightarrow$ Hyper-Kamiokande 
experiment and a few years running of a reactor experiment with 
100ton detectors at the Kashiwazaki-Kariwa nuclear power plant.
We show that the method can be arranged to be insensitive 
to the intrinsic parameter degeneracy but is affected by 
the one due to unknown sign of $\Delta m^2_{31}$. 

\end{abstract}


\maketitle


\section{Introduction}

After the pioneering and the long-term extensive efforts in the 
atmospheric \cite{SKatm}, the solar \cite{solar}, the accelerator 
\cite{K2K}, and the reactor \cite{KamLAND} experiments, 
we have grasped the structure of lepton flavor mixing in 
the (2-3) and the (1-2) sectors of the Maki-Nakagawa-Sakata (MNS) 
matrix \cite{MNS}. 
Now we are left with the unique unknown (1-3) sector of the 
MNS matrix, in which there live the third mixing angle 
$\theta_{13}$, which is known to be small \cite{CHOOZ}, and 
the completely unknown CP-violating leptonic Kobayashi-Maskawa 
phase $\delta$ \cite{KM}.

Detecting leptonic CP violation is one of the most challenging 
goals in particle physics.
A popular method for measuring the 
CP-violating phase is by long-baseline (LBL) accelerator neutrino 
experiments using either conventional neutrino superbeam 
\cite{J-PARC,SPL,BNL}, or an intense beam from muon storage ring 
\cite{nufact}. If $\theta_{13}$ is not too small, it is likely that 
leptonic CP violation is first explored by LBL experiments with 
conventional superbeam \cite{superbeam}. 

To measure CP-violating phase the LBL experiments must run 
not only with the neutrino mode but also with the antineutrino mode.
Apart from the problem of parameter degeneracy 
\cite{Burguet-C,MNjhep01,KMN02,octant,BMW1,MNP2}, 
these measurement would allow us 
to determine the CP-violating phase $\delta$ to a certain accuracy. 
In the Japan Proton Accelerator Research Complex (J-PARC) 
$\rightarrow$ Hyper-Kamiokande project with upgraded 
4MW beam of 50GeV accelerator at J-PARC, the accuracy of 
determination of $\delta$ is expected to be $\simeq$ 20 degrees at 
3$\sigma$~CL \cite{J-PARC}.

Running the experiment with antineutrino mode, however, is 
possible only by overcoming a variety of difficulties, 
much greater ones compared with those in neutrino mode operation. 
Even if we ignore the issue of slightly less intense 
$\pi^{-}$ beam compared to $\pi^{+}$ beam, the antineutrino 
cross sections are smaller by factor of $\simeq$ 3 than neutrino cross 
sections, which results in three-times longer period of data taking, 
6 years of $\bar{\nu}$-mode compared to 2 years of $\nu$-mode 
operation in the J-PARC $\rightarrow$ Hyper-Kamiokande 
(hereafter abbreviated as JPARC-HK) experiment. Moreover, 
the background in $\bar{\nu}_{e}$ appearance detection, 
according to the current estimate, are larger by factor of 
$\simeq$~2 compared with those in $\nu_{e}$ detection. 
Hence, antineutrino-mode measurement may be better characterized as 
an independent experiment rather than the in-situ measurement. 
Considering three times longer running time it is certainly 
worthwhile to think about an alternative which can run 
simultaneously with neutrino-mode superbeam appearance experiments.

In this paper, we point out that a reactor experiment can serve for 
such purpose. We demonstrate that reactor experiments for 
measuring $\theta_{13}$ with reasonable assumptions on their 
systematic errors can uncover the leptonic CP violation 
when combined with high-statistics neutrino-mode superbeam 
experiments.
In fact, we have pointed out such possibility in our previous 
communication \cite{MSYIS}, in which we have demonstrated the 
complementary role of the reactor and the LBL accelerator 
experiments in determination of the remaining neutrino 
mixing parameters.
The treatment in this paper quantifies our proposal and 
thereby complements and further strengthen our viewpoint of the 
LBL-reactor complementarity.
A quantitative treatment of sensitivity for detecting 
CP violation by reactor-LBL combination was also attempted in 
Ref.~\cite{Munich} but with no indication of signal. 
See Ref.~\cite{krasnoyarsk,kashkari,diablo} for detailed 
description of possible experimental designs for reactor 
experiments for measuring $\theta_{13}$.

We remark that the sensitivity to CP violation by our 
reactor-LBL combined method suffers from the problem of 
parameter degeneracy. However, it can be arranged so that 
it is insensitive to the intrinsic parameter degeneracy 
\cite{Burguet-C}.
If the superbeam experiment is done at the oscillation maximum 
the combined measurement will allow us to determine 
$\sin{\delta}$. 
Obviously, the measurement by itself cannot resolve the 
ambiguity $\delta \leftrightarrow \pi - \delta$, 
but it does not produce a fake CP violation. 
We have to note that our method suffers from the problem of 
degeneracy due to unknown sign of $\Delta m^2_{31}$ \cite{MNjhep01}.
Even in the case of the JPARC-HK experiment in which the 
matter effect is only modest, it does affects the CP 
sensitivity because the degenerate solutions of $\delta$ 
differ by $\sim \pi/2$ in overlapping region of two 
ellipses in the bi-probability plot \cite{MNjhep01}. 
Therefore, it is important to know the sign of $\Delta m^2_{31}$ 
prior to the reactor-LBL measurement of $\delta$.
While the octant ambiguity of $\theta_{23}$ \cite{octant} 
may also affect the CP sensitivity, we do not try to elaborate 
this point in the present paper. 
See, however, (1) in the concluding remarks.

We emphasize that reactor experiment cannot replace 
the antineutrino-mode superbeam experiments. It is because  
the reactor-LBL combined method can detect leptonic CP 
violation only up to $\simeq$~2$\sigma$~CL\@. Nevertheless, we believe 
that such reactor-LBL combined measurement has a great merit. 
It will give us the first grip of the structure of leptonic CP violation. 
It will also merit the then ongoing neutrino-mode and the 
following antineutrino-mode superbeam experiments themselves; 
Even a rough knowledge of the feature of 
CP violation would be very helpful to optimize the setting 
(such as relative time sharing of $\nu$ and $\bar{\nu}$ modes) 
of the difficult and extremely long-term experiment.

\section{Reactor-LBL Combined Measurement of CP Violation}

The principle of detection of leptonic CP violation in a 
reactor-LBL combined measurement is very simple.
First let us remind the readers the characteristic features of 
reactor measurement of $\theta_{13}$. 
As we have discussed in length in Ref.~\cite{MSYIS}, 
reactor experiment can serve for pure measurement of $\theta_{13}$ 
assuming that $\Delta m^2_{31}$ is accurately determined by 
disappearance measurement of $P(\nu_{\mu} \rightarrow \nu_{\mu})$ 
in LBL experiments. 
Namely, it is not contaminated by uncertainties due to unknown 
CP phase $\delta$, the matter effect, and possibly to the 
octant ambiguity $\theta_{23} \rightarrow \pi/2 - \theta_{23}$ 
from which $\nu_{e}$ appearance measurement by 
LBL experiment suffers. 

Now LBL $\nu_{e}$ appearance experiment will observe the 
neutrino oscillation probability $P(\nu_{\mu} \rightarrow \nu_{e})$. 
In leading order in $\Delta m^2_{21}/\Delta m^2_{31}$ 
it takes the form 
\cite{golden}
\begin{eqnarray}
P(\nu_{\mu} \rightarrow \nu_{e}) \equiv 
P(\nu)_{\pm} = X_{\pm} s_{13}^2 + 
Y_{\pm} s_{13} \cos {\left( \delta \pm \frac{\Delta_{31}}{2} \right)} + 
P_{\odot},
\label{Pmue}
\end{eqnarray}
where $\pm$ refers to the sign of $\Delta m^2_{31}$. 
The coefficients $X_{\pm}$,  $Y_{\pm}$, and $P_{\odot}$ 
are given by 
\begin{eqnarray}
X_{\pm} &=& 4 s^2_{23} 
\left(
\frac{\Delta_{31}}{B_{\mp}} 
\right)^2
\sin^2{\left(\frac{B_{\mp}}{2}\right)}, 
\label{X} \\
Y_{\pm} &=& \pm 8 c_{12}s_{12}c_{23}s_{23}
\left(
\frac{\Delta_{21}}{aL}
\right)
\left(
\frac{\Delta_{31}}{B_{\mp}}
\right)
\sin{\left(\frac{aL}{2}\right)}
\sin{\left(\frac{B_{\mp}}{2}\right)},
\label{Y}\\
P_{\odot} & = & c^2_{23} \sin^2{2\theta_{12}} 
\left(\frac{\Delta_{21}}{aL}\right)^2
\sin^2{\left(\frac{aL}{2}\right)}
\end{eqnarray}
with 
\begin{eqnarray}
\Delta_{ij}  \equiv \frac{|\Delta m^2_{ij}| L}{2E}
\quad
{\rm and} \quad B_{\pm} \equiv \Delta_{31} \pm aL,
\end{eqnarray}
where $a = \sqrt{2} G_F N_e$ denotes the index of refraction 
in matter with $G_F$ being the Fermi constant and $N_e$ a constant 
electron number density in the earth. 
We use in this paper the standard notation of the MNS matrix 
\cite{PDG}.
The mass squared difference of neutrinos 
is defined as $\Delta m^2_{ji} \equiv m^2_j - m^2_i$ where
$m_i$ is the mass of the $i$-th eigenstate.

There exist number of reasons for tuning the beam energy to 
the oscillation maximum $\Delta_{13} = \pi$ in doing the appearance 
and the disappearance measurement in LBL experiments, 
as listed in \cite{KMN02}. In this case, 
$\cos{\left(\delta \pm \frac{\Delta_{13}}{2} \right)} = \mp \sin{\delta}$ 
and (\ref{Pmue}) can be solved for  $\sin{\delta}$ as 
\begin{eqnarray}
\sin{\delta} = 
\frac{P(\nu) - P_{\odot} - X_{\pm} s_{13}^2}
{\mp Y_{\pm} s_{13}}.
\label{sdelta}
\end{eqnarray}
We note that, since $\theta_{13}$ can be measured by reactor 
experiments, the right-hand side (RHS) of (\ref{sdelta}) consists 
solely of experimentally measurable quantities. Therefore, LBL 
measurement of $P(\nu_{\mu} \rightarrow \nu_{e})$, when combined 
with the reactor experiment, implies measurement of $\sin{\delta}$.

In the rest of this paper, we try to elaborate our treatment by 
including suitably estimated experimental uncertainties of both 
LBL and the reactor experiments. 
As indicated in (\ref{sdelta}), the accuracy of measurement of 
$\sin{\delta}$ solely depends upon how precisely $P(\nu)$ and 
$s_{13}$ in the RHS can be determined in LBL and reactor 
experiments, respectively. 
We take the best possible case among the concrete proposals of 
LBL experiments currently available in the community,  
the JPARC-HK experiment assuming 4MW beam power and 540 kton as 
the fiducial volume of the detector \cite{Hyper-K}.
However, most probably our conclusion does not heavily depend 
on any detailed experimental setting in the particular experiment, 
once the accuracy of measurement of the $\nu_{e}$ appearance 
probability reaches to that level and if the baseline is not too long.
For the reactor experiment, we present our results in units 
of $\GWth\cdot\mbox{ton}\cdot\mbox{year}$ exposure to allow 
application to wider class of experiments.
Our results may be useful to indicate what condition must be met 
to uncover the leptonic CP violation in such reactor-LBL combined 
measurement.

\section{Treatment of errors in LBL and reactor experiments}

To carry out quantitative analyses of the sensitivity for detecting 
CP violation, we must first establish the method for 
statistical treatment of LBL and reactor experiments.

\subsection{Treatment of errors in the JPARC-HK experiment}

We consider neutrino-mode appearance measurement for 2 years in 
the JPARC-HK experiment. For definiteness, we use the neutrino flux 
estimated for the off-axis $2^\circ$ beam \cite{J-PARC}.
We define $\Delta\chi^2$ for the experiment as 
\begin{eqnarray}
 \Delta\chi^2_{\J-PARC\nu}
 \equiv \frac{ \left( N_\nu - N_\nu^\best \right)^2 }
       { N_\nu^\best + N_\BG + \sigS^2 (N_\nu^\best)^2 + \sigBG^2 (N_\BG)^2 },
\end{eqnarray}
where $N_\nu$ and $N_\BG$ represent the expected number of signal
and background events, respectively, computed with the cross section 
in \cite{kameda}.
$N_\nu^\best$ is defined as the number of signal event $N_\nu$ 
calculated with the best-fit values of the ``experimental data'', 
which is to be tested against the CP conserving hypothesis, 
$\delta = 0$.
$\sigS$ and $\sigBG$ represent the fractional uncertainties of 
the estimation of the number of signal and background events, 
respectively. Following \cite{J-PARC}, we use 
$\sigS = \sigBG = 2\%$ in our analysis.
(See Sect.~\ref{sectCPV} for more about how to use $\Delta \chi^2$ in 
our procedure to determine the sensitivity region
for CP violation.)

While we do not use the spectrum information in a direct way in 
our analysis, we need to estimate how the experimental event 
selection affects the spectrum to calculate the number of signal 
and background events. 
The most important cut is to suppress the background events due to 
$\pi^0$.
We use the simulated spectra after the cut calculated by the 
JPARC-SK group \cite{kobayashi} and evaluate the reduction rate 
due to cut in each energy bin of 50MeV width. 
The procedure is applied to calculate the number of signal 
after the cut for any values of mixing parameters. 
In this way, the total numbers of events within energy 
range \mbox{0.4-1.2GeV} are calculated and used in our analysis.

\subsection{Treatment of errors in the reactor experiment}

In this paper we consider the case of single reactor and 
two (near and far) detector complex.\footnote{
The current proposal by the Japanese group \cite{kashkari}
plans to utilize $\bar{\nu}_e$ flux from 7 reactors observed 
by two near and a far detectors. It is shown even in this 
case that an effective 1 reactor-2 detector approximation gives a 
very good estimation of the sensitivity \cite{SSY03}.
}
The far detector is placed 1.7km away from the reactor, 
the optimal distance for 
$|\Delta m^2_{31}| = 2.5\times 10^{-3}\,\eV^2$.
We assume that a near detector identical with the 
far detector is placed 300m away from the reactor 
to reduce systematic errors.\footnote{
The closer the near detector to reactor, the better 
the sensitivity in the single-reactor case because of 
larger number of events. The situation is, 
however, more subtle for multiple-reactor case \cite{SSY03}.
}

We consider four types of systematic errors:
$\sigDB$, $\sigDb$, $\sigdB$, and $\sigdb$.
The subscript D (d) represents the fact that the error is 
correlated (uncorrelated) between detectors.
The subscript B (b) represents that the error is correlated 
(uncorrelated) among bins.
To indicate nature of these respective errors, we list below 
some examples of the errors in each category:

\noindent \hskip 1.0cm
$\sigDB$: error in estimation of reactor power

\noindent \hskip 1.0cm
$\sigDb$: error in estimation of detection cross sections

\noindent \hskip 1.0cm
$\sigdB$: error in estimation of fiducial volume of each detector 

\noindent \hskip 1.0cm
$\sigdb$: errors inherent to detectors such as artificial firing of 
photomultiplier tubes 

\noindent 
Although the values of $\sigdB$ for far and near detectors, for example,
can be different from each other,
we neglect such difference for simplicity.
The values of systematic errors we assume are listed
in Table~\ref{syserror}.

\begin{table}
\begin{tabular}{|c|c|c|c||c|}
 \hline
 \multicolumn{2}{|c|}{} & \multicolumn{2}{|c||}{between detectors} & \\
 \cline{3-4}
 \multicolumn{2}{|c|}{} & correlated & uncorrelated & \ single detector \ \\
 \hline
 \ between bins \ & correlated & \ $\sigDB = 2.5\%$ \ & \
 $\sigdB = 0.5\%$ \ & $\sigB \simeq 2.6\%$ \\
 \cline{2-5}
 & \ uncorrelated \ & $\sigDb = 2.5\%$ &
 $\sigdb = 0.5\%$ & $\sigb \simeq 2.6\%$ \\
 \hline\hline
 \multicolumn{2}{|c|}{ \ total number of events \ } &
 $\sigD \simeq 2.6\%$ &
 $\sigd \simeq 0.5\%$ &
 \ $ \sigma_\sys \simeq 2.7\%$ \ \\
 \hline
\end{tabular}
\caption{Listed are assumed values of systematic errors
$\sigDB$, $\sigDb$, $\sigdB$, and $\sigdb$.
 The subscripts D (d) and B (b) are represent
the correlated (uncorrelated) error among detectors and bins,
respectively.
 Using those four values,
the errors for the total number of events and for single detector
are calculated.}
\label{syserror}
\end{table}

As will be briefly explained in the appendix, the errors 
$\sigD$ and $\sigd$ for the total number of events are obtained as
\begin{eqnarray}
 \sigD^2 = \sigDB^2 + \sigDb^2 \frac{ \sum_i (N_{ai}^\best)^2 }
                                    { \left( \sum_i N_{ai}^\best \right)^2 }, \ \
 \sigd^2 = \sigdB^2 + \sigdb^2 \frac{ \sum_i (N_{ai}^\best)^2 }
                                    { \left( \sum_i N_{ai}^\best \right)^2 },
\label{error_rel}
\end{eqnarray}
where $a=n, f$ are the index for near and far detectors, 
and $i$ runs over number of bins.
 We use 14 bins of $0.5 \MeV$ width in 1-8MeV window of visible energy,
$E_{\text{visi}} = E_{\bar{\nu}_e} - 0.8 \MeV$.
The coefficient of $\sigDb^2$ and $\sigdb^2$ is
about 1/9 in our analysis almost independently of $a$.
Since relative normalization errors are $\sqrt{2}$ times
of uncorrelated errors, $\sigd \simeq 0.5\% $ is
consistent with the value used in \cite{MSYIS}.
In Ref.~\cite{MSYIS}, the most pessimistic assumption 
$\sigDB = \sigdB = 0$ was taken for bin-by-bin distribution of errors.
The value of $\sigma_\sys^2 \equiv \sigD^2 + \sigd^2$ is
also consistent with the total systematic error of the CHOOZ experiment.
In summary, we feel that the errors listed in Table~\ref{syserror}
are not too optimistic ones and are likely to be realized in 
the setting discussed in \cite{krasnoyarsk,kashkari,diablo}.

 Our definition of $\Delta\chi^2_\react$ is
\begin{eqnarray}
 \Delta\chi^2_\react
 &\equiv&
 \min_{\text{$\alpha$'s}}
  \sum_{a = f, n}
          \left[
           \sum_{i=1}^{14}
            \left\{
	     \frac{
	      \left( N_{ai}
	             - ( 1 + \alpha_i + \alpha_a + \alpha ) N^\best_{ai}
	      \right)^2 }
	      { N^\best_{ai}
                + \sigdb^2 (N^\best_{ai})^2 }
             + \frac{ \alpha_i^2 }{ \sigDb^2 }
	    \right\}
	   + \frac{ \alpha_a^2 }{ \sigdB^2 }
	  \right]
	  + \frac{ \alpha^2 }{ \sigDB^2 },
\label{chireac}
\end{eqnarray}
where $N_{ai}$ represents the theoretical number of events
at $a$-detector within $i$-th bin. 
Again, $N_{ai}^\best$ is defined as the number of signal event 
calculated with the best-fit parameters of the ``experimental data''.  
The minimization in (\ref{chireac}) is achieved analytically,
and then we obtain
\begin{eqnarray}
 &&
\Delta\chi^2_\react
 = ( \vec{x}^{\text{T}}, \vec{y}^{\text{T}} ) V^{-1}
         \left(
          \begin{array}{c}
	   \vec{x} \\
	   \vec{y} \\
	  \end{array}
	 \right),\\[5mm]
&&
\vec{x}^{\text{T}}
 \equiv \left(
	 \frac{ N_{f1} - N^\best_{f1} }{ N^\best_{f1} },
	 \cdots
	\right), \ 
\vec{y}^{\text{T}}
 \equiv \left(
	 \frac{ N_{n1} - N^\best_{n1} }{ N^\best_{n1} },
	 \cdots
	\right),\\
&&
V \equiv \text{diag}\left(
		     \frac{1}{N^\best_{f1}}, \cdots,
		     \frac{1}{N^\best_{n1}}, \cdots
		    \right)\nonumber\\
&&\hspace{20mm}
       {}+ \sigdb^2 I_{28}
	 + \sigdB^2 \left(
		     \begin{array}{cc}
		      H_{14} & 0 \\
		      0 & H_{14}
		     \end{array}
		    \right)
	 + \sigDb^2 \left(
		     \begin{array}{cc}
		      I_{14} & I_{14} \\
		      I_{14} & I_{14}
		     \end{array}
		    \right)
	 + \sigDB^2 H_{28},
\end{eqnarray}
where $I_n$ represents the $n\times n$ identity matrix
and $H_n$ represents the $n\times n$ matrix
whose elements are all unity.
Notice that an infinitely good sensitivity is obtained
for infinite number of events if $\sigdb$ vanishes 
because $\det (V)$ goes to zero for the case which explains 
the apparently curious behavior seen in Fig.~2 of \cite{Munich}.
See \cite{stump} for more about the equivalence between the 
``pull'' and the covariance matrix methods.

\begin{figure}[htbp]
\begin{center}
\includegraphics[width=0.7\textwidth]{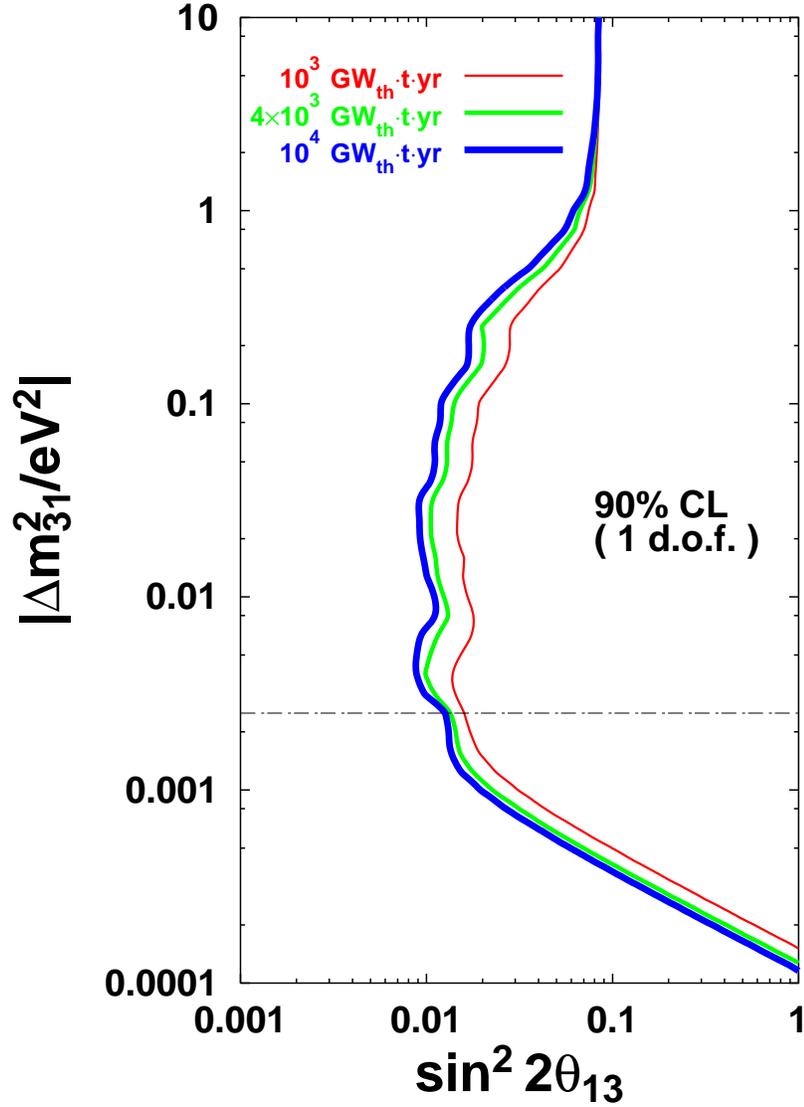}
\end{center}
\caption{The excluded regions at 90\%~CL in the absence of 
$\bar{\nu}_e$ disappearance ($\theta_{13}^\best = 0$) are 
drawn for $10^3$, $4\times 10^3$, and $10^4$ 
$\GWth\cdot\mbox{ton}\cdot\mbox{year}$ exposure of a reactor experiment 
by thin-solid (red), solid (green), and thick-solid (blue) lines, 
respectively.
The far (near) detector is placed 1.7km (300m) away from the reactor.
We assume that $|\Delta m^2_{31}|$ is precisely measured by LBL 
experiments and adopt the analysis with one degree of freedom 
($\Delta\chi^2_\react = 2.7$).
We use the value $|\Delta m^2_{31}| = 2.5\times 10^{-3} \eV^2$
as indicated by the dashed-doted line in the figure.
In our analysis, we use 14~bins of 0.5MeV width in \mbox{1-8MeV} 
window of visible energy with the systematic errors listed 
in Table~\ref{syserror}.
}
\label{th13}
\end{figure}

To indicate the expected sensitivity of the reactor experiment 
with the systematic errors listed in Table~\ref{syserror}, 
we present in Fig.~\ref{th13} the excluded region in 
$\sin^2{2\theta_{13}}$-$|\Delta m^2_{31}|$ space 
in the absence of flux depletion ($\theta_{13}^\best = 0$) for 
$10^3$, $4 \times 10^3$, and $10^4$ 
$\GWth\cdot\mbox{ton}\cdot\mbox{year}$ exposure. 
The $\bar{\nu}_e$ detection efficiency of 70\% is assumed 
\cite{CHOOZ,MSYIS}. 
The number of events expected during these exposure are 
about $10^5$, $4\times10^5$, $10^6$ $\bar{\nu}_e$ events, 
respectively, at the far detector.\footnote{
In the rate-only analysis without binning, 
the sensitivity is saturated at the number of $\bar{\nu}_e$ events 
around $10^5$.
}
Notice that what we mean by numbers in units of $\GWth$ is  
the thermal power actually generated from reactors and 
it should not be confused with the maximal 
thermal power of reactors.
Assuming average 80\% operation efficiency 
the above three cases correspond approximately to 
0.5, 2, and 5 years running, respectively, for 100~ton detector 
at the Kashiwazaki-Kariwa nuclear power plant whose maximal 
thermal power is $24.3\GWth$.

\section{Estimation of Sensitivity of Reactor-LBL Combined 
Detection of CP Violation\label{sectCPV}}

To estimate the sensitivity of the reactor-LBL combined 
measurement to leptonic CP violation, we define the combined 
$\Delta\chi^2$ as 
\begin{eqnarray}
 \Delta\chi^2_\CP1 (\delta; \delta^\best, \sin^2{2\theta_{13}^\best})
 &\equiv&
 \min_{\sin^2{2\theta_{13}}}
 \Delta\chi^2_\rCP (\delta, \sin^2{2\theta_{13}}; \delta^\best, \sin^2{2\theta_{13}^\best}) \nonumber\\
 &\equiv&
 \min_{\sin^2{2\theta_{13}}}
 \Bigl\{
  \Delta\chi^2_{\J-PARC\nu}(\delta, \sin^2{2\theta_{13}}; \delta^\best, \sin^2{2\theta_{13}^\best}) \nonumber\\
 &&
\hspace*{35mm}
 {}+ \Delta\chi^2_\react(\sin^2{2\theta_{13}}; \sin^2{2\theta_{13}^\best})
 \Bigr\}.
\label{combine}
\end{eqnarray}
We take the following procedure in our analysis.
We pick up a point in the two-dimensional parameter space 
spanned by $\delta^\best$ and $\sin^2{2\theta_{13}^\best}$ 
and make the hypothesis test on whether the point is consistent 
with CP conservation within 90\%~CL. For this purpose, 
we use the projected $\Delta\chi^2$ onto one-dimensional 
$\delta$ space, $\Delta\chi^2_\CP1$, as defined in 
(\ref{combine}) and then the statistical criterion for 
90\%~CL is $\Delta\chi^2_\CP1 \leq 2.7$.
Then, a collection of points in the parameter space which 
are consistent with CP conservation form a region surrounded 
by a contour in 
$\delta^\best$-$\sin^2{2\theta_{13}^\best}$ space, 
as will be shown in \mbox{Figs.~\ref{CPn}-\ref{CPd}} below.

The neutrino mixing parameters are taken as follows:
$|\Delta m_{31}^2| = 2.5\times10^{-3}\eV^2$,
$\Delta m_{21}^2 = 7.3\times10^{-5}\eV^2$,
$\tan^2{\theta_{12}} = 0.38$,
and $\sin^2{2\theta_{23}} = 1$.
Notice that the high-$\Delta m^2_{21}$ LMA-II solar neutrino 
solution is now excluded at 3$\sigma$~CL by the global analysis of 
all data with reanalyzed day-night variation of flux at 
Super-Kamiokande \cite{SK_daynight}, and at 
99\%~CL by the one with SNO salt phase data \cite{SNO_salt}.
The earth matter density is
taken to be $\rho = 2.3\,\mbox{g}\cdot\mbox{cm}^{-3}$ 
\cite{koike-sato} and the electron number density is 
computed with electron fraction $Y_e=0.5$.

\subsection{CP sensitivity in the case of known sign of $\Delta m^2_{31}$}

\begin{figure}[htbp]
\begin{center}
\includegraphics[width=0.9\textwidth]{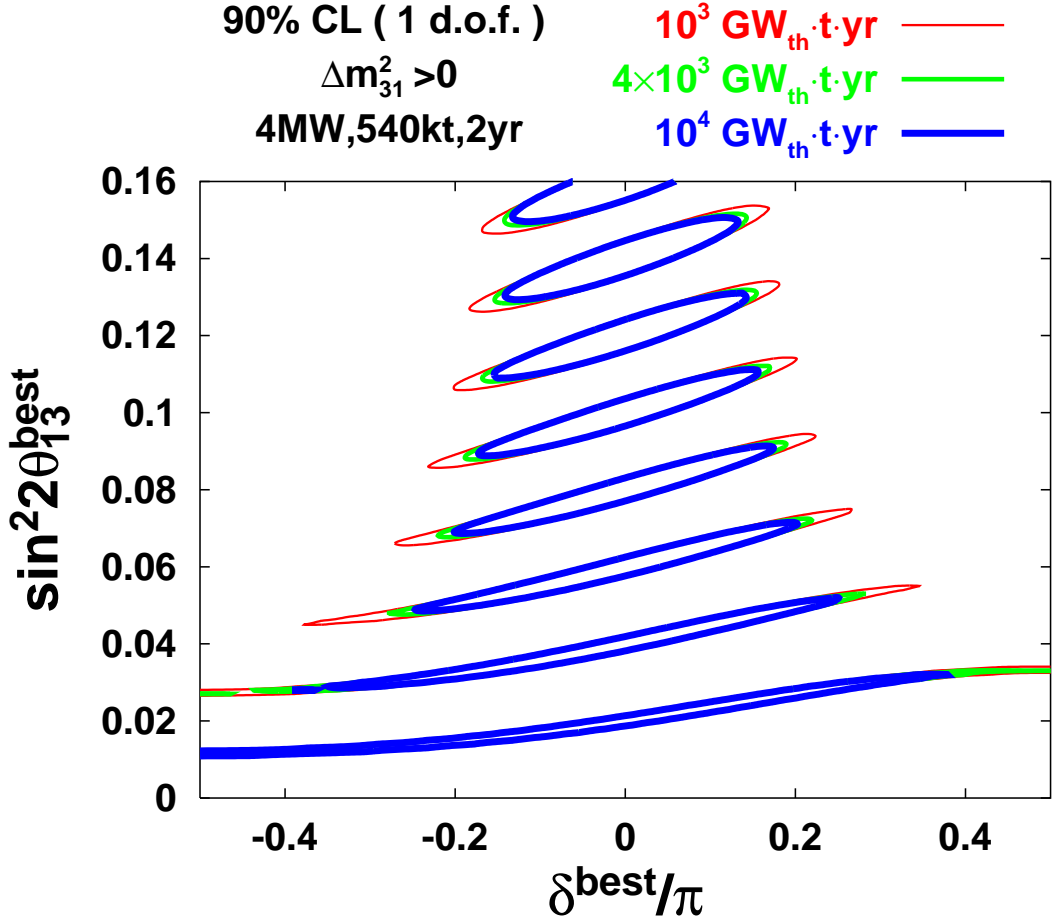}
\end{center}
\caption{The contours are plotted for eight assumed values of 
$\sin^2{2\theta_{13}}$ which range from 0.02 to 0.16 
to indicate the regions consistent with the hypothesis 
$\delta = 0$ at 90\%~CL ($\Delta\chi^2_\rCP = 2.7$) 
by the reactor-LBL combined measurement.
If an experimental best fit point falls into outside the envelope of 
those regions, it gives an evidence for leptonic CP violation at 
90\%~CL\@. 
The thin-solid (red), solid (green), and thick-solid (blue) 
lines are for 
$10^3$, $4\times 10^3$, and $10^4 \GWth\cdot\mbox{ton}\cdot\mbox{year}$ 
exposure of a reactor experiment, respectively, 
corresponding to about 0.5, 2, and 5~years exposure of 
100~ton detectors at the Kashiwazaki-Kariwa nuclear power plant.
For the JPARC-HK experiment, 2~years measurement with 
off-axis $2^\circ$ $\nu_\mu$ beam is assumed. 
(See the text for more details.)
The normal mass hierarchy, $\Delta m^2_{31} > 0$, is assumed.
}
\label{CPn}
\end{figure}

In Fig.~\ref{CPn}, the regions consistent with CP conservation 
at 90\%~CL are drawn for $\Delta m^2_{31} > 0$ case in the region 
$- \pi/2 \leq \delta^\best \leq \pi/2$. 
The thin-solid, the solid, and the thick-solid lines are 
for $10^{3}$, $4 \times 10^{3}$, and
$10^{4}\GWth\cdot\mbox{ton}\cdot\mbox{years}$, respectively,  
and the regions consistent with CP conservation are within 
the envelope of these contours.\footnote{
Since we rely on hypothesis test with 1 degree of freedom (1 d.o.f.) 
the information of $\sin^2{2\theta_{13}}$ is lost through 
the process of minimization in (\ref{combine}). 
The individual contours presented in Fig.~\ref{CPn} indicate 
the region $\Delta \chi^2_\rCP \leq 2.7$ for eight assumed 
values of $\sin^2{2\theta_{13}}$ which range from 0.02 to 0.16. 
In this way the figure is designed so that the envelop of 
the contours gives the region of CP conservation at 90\%~CL 
by 1 d.o.f.\ analysis, 
and at the same time carries some informations of how the 
sensitivity regions are determined by the interplay between 
the reactor and the LBL measurement.
We hope that no confusion arise.
}
We remark that the present constraint on $\theta_{13}$ becomes milder 
to $\sin^2{2\theta_{13}} < 0.25$ at 3$\sigma$~CL \cite{bari_update} 
by the smaller values of $|\Delta m^2_{31}|$ indicated by the 
reanalysis of atmospheric neutrino data \cite{hayato}.
Notice that the other half region of $\delta^\best$ gives the 
identical contours apart from tiny difference which arises 
because the peak energy of the off-axis $2^\circ$ 
beam is slightly off the oscillation maximum.

If an experimental best fit point falls into outside the envelope 
of those regions, it gives an indication for leptonic CP violation
because it is inconsistent with the hypothesis $\delta = 0$
at 90\%~CL\@.
We observe from Fig.~\ref{CPn} that there is a chance for 
reactor-LBL combined experiment of seeing an indication of CP 
violation for relatively large $\theta_{13}^\best$, 
$\sin^2{2\theta_{13}^\best} \geq 0.03$ at 90\%~CL\@.
We believe that this is the first time that a possibility is 
raised for detecting leptonic CP violation 
based on a quantitative treatment of experimental errors by 
a method different from the conventional one of comparing 
neutrino and antineutrino appearance measurement in LBL experiments.

The sign of $\Delta m^2_{31}$ is taken to be positive in 
Fig.~\ref{CPn} which corresponds to the normal mass hierarchy.
If we flip the sign of $\Delta m^2_{31}$ 
(the case of inverted mass hierarchy) 
we obtain almost identical CP sensitivity contours.
It is demonstrated in Figs.~\ref{CPd}a and \ref{CPd}b which serve 
also for the discussion in the next subsection. By comparing 
the contours depicted by thick-solid and thick-dashed lines in 
Figs.~\ref{CPd}a ($\Delta m^2_{31} > 0$) and 
\ref{CPd}b ($\Delta m^2_{31} < 0$), respectively, it is clear 
that the CP sensitivity is almost identical between positive 
and negative $\Delta m^2_{31}$. 
The largest noticeable changes are shifts of the end points 
of the contours toward smaller (larger) $\delta$ in the 
first (fourth) quadrants by about 10\% (a few \%) 
at $\sin^2{2\theta_{13}} = 0.1$.
Namely, the both end points slightly move toward better sensitivities
for the inverted mass hierarchy.

The sensitivity contour of CP violation is determined as an 
interplay between constraints from reactor and accelerator 
experiments. The former gives a rectangular box in the 
$\delta^\best$-$\sin^2{2\theta_{13}^\best}$ space, whereas the 
latter gives the equal-$P(\nu)$ contour determined by 
(\ref{Pmue}) under the hypothesis $\delta=0$ 
with finite width due to errors, as indicated
in Figs.~\ref{CPn} and \ref{CPd}.
In region of parameter space where both of these two constraints 
are satisfied, the best fit parameter is consistent with CP conservation.
Outside the region the CP symmetry is violated at 90\%~CL\@.
The discovery potential for CP violation diminishes at small 
$\sin^2{2\theta_{13}^\best}$ primarily because $P(\nu)$ becomes 
less sensitive to $\delta$ at smaller $\theta_{13}$, 
while the reactor constraint on $\sin^2{2\theta_{13}^\best}$ 
is roughly independent of $\theta_{13}$ \cite{MSYIS}.

\subsection{CP sensitivity in the case of unknown sign of $\Delta m^2_{31}$}

So far we have assumed that we know the sign of 
$\Delta m^2_{31}$ prior to the search for CP violation by the 
reactor and the JPARC-HK experiments. But, it may not be the 
case unless LBL experiments with sufficiently long baseline 
start to operate in a timely fashion. 
In this subsection we assume the pessimistic situation of 
unknown sign of $\Delta m^2_{31}$ and try to clarify the 
influence of our ignorance of the sign on the detectability 
of CP violation by our method. 

\begin{figure}[htbp]
\includegraphics[width=0.48\textwidth]{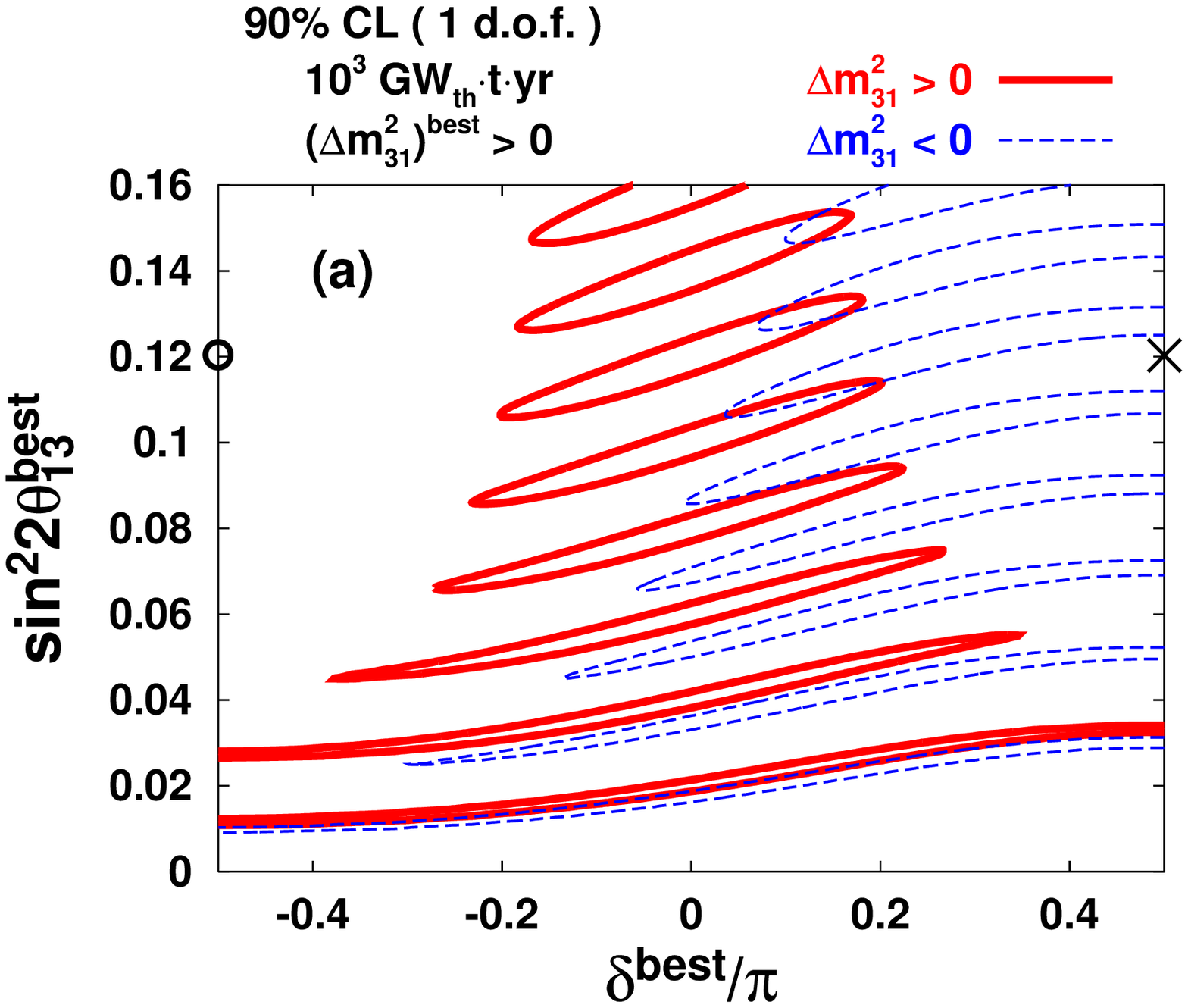}
\includegraphics[width=0.48\textwidth]{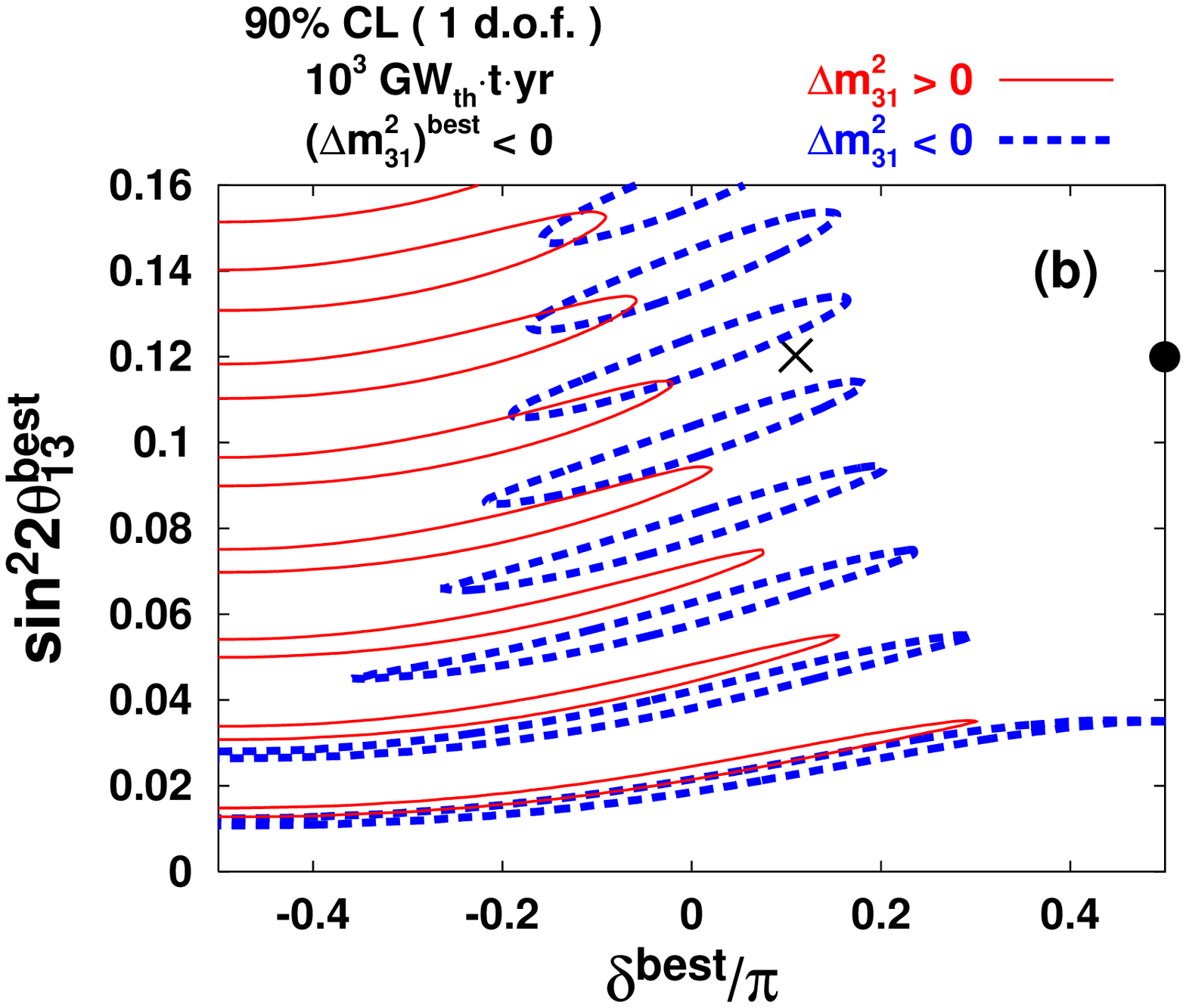}
\caption{
The contours which surround the region consistent with 
CP conservation are plotted in Fig.~\ref{CPd}a~(\ref{CPd}b) by assuming 
$(\Delta m^2_{31})^\best > 0$ ($(\Delta m^2_{31})^\best < 0$)
as nature's choice. 
If the right (wrong) sign is used as the hypothesis with $\delta = 0$,
the contours indicated by the thick (thin) lines result in both figures. 
The three symbols, a cross, open and solid circles are 
placed on the figures as well as in Fig.~\ref{CPd-eve} to indicate 
the relationship between observed numbers of events and 
the results of CP sensitivity analysis. 
}
\label{CPd}
\end{figure}

If the sign of $\Delta m^2_{31}$ is not known, the procedure 
of obtaining the sensitivity region for detecting CP violation 
has to be altered. It is because we have to allow such possibility 
as that we fit the data by using wrong assumption for the sign. 
In Fig.~\ref{CPd}a~(\ref{CPd}b) we present the results of the similar sensitivity 
analysis for detecting CP violation as we did in the previous 
section by assuming that the sign of $(\Delta m^2_{31})^\best$,
which is chosen by nature, is positive (negative).
 It is obvious from Fig.~\ref{CPd}a~(\ref{CPd}b) that the contours of 
CP conservation moves to rightward (leftward)
if the wrong sign is assumed in the hypothesis test,
essentially wiping out 
about half of the CP sensitive region of $\delta^\best$.

\begin{figure}[htbp]
\includegraphics[width=0.7\textwidth]{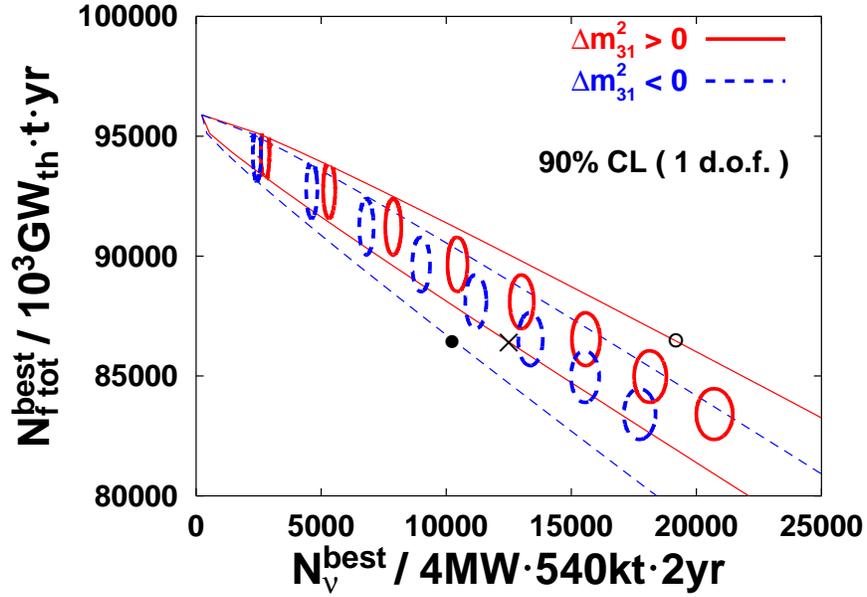}
\caption{
The contours which surround the region consistent with 
CP conservation are plotted in the number-of-events space 
of reactor and the LBL experiments. 
The thin lines correspond to $\delta^\best = \pm \pi/2$.
The three symbols, a cross, open and solid circles are 
placed on the figures as well as in Fig.~\ref{CPd} to indicate 
the relationship between observed numbers of events and 
the results of CP sensitivity analysis.
}
\label{CPd-eve}
\end{figure}

The results can be confusing and some of the readers might 
have naively interpreted, by combining Figs.~\ref{CPd}a and \ref{CPd}b, that 
there is no sensitivity region
in $\delta^\best\mbox{-}\sin^2{2\theta_{23}^\best}$ plane. 
To resolve the puzzling feature we present in Fig.~\ref{CPd-eve} the regions 
which are consistent with CP conservation by contours in the plane 
spanned by observable quantities, the number of events in the 
reactor and the JPARC-HK $\nu_e$ appearance experiments.\footnote{
Notice, however, that we have used binned data, not merely the 
total number of events, in analyzing reactor experiment to 
obtain the contours.
}
This plot indicates that the sensitivity region for detecting 
CP violation does not disappear but becomes about half. 
Which region of $\delta$ is CP sensitive depend upon the 
sign of $\Delta m^2_{13}$, or in other word 
on the location in bi-number of event plane in Fig.~\ref{CPd-eve}.
For complete clarity, we have placed three different symbols in 
Fig.~\ref{CPd-eve} and at the same time in Fig.~\ref{CPd} to indicate 
which points in the space of observables correspond to 
which points in the CP sensitivity plot.
 Note that the point indicated by a cross in Fig.~\ref{CPd-eve} 
corresponds to two values of $\delta^\best$
because of unknown sign of $\Delta m^2_{13}$.

\section{Concluding remarks}

In this paper we have pointed out a new method for detecting 
leptonic CP violation by combining reactor measurement of 
$\theta_{13}$ with high-statistics $\nu_e$ appearance 
measurement in LBL accelerator experiments. 
A salient feature of our method are that one can perform the 
measurement prior to the lengthy antineutrino running 
in LBL experiments.
We conclude with several remarks:

\vskip 0.3cm
\noindent
(1) If $\theta_{23}$ is not maximal the parameter degeneracy 
due to the octant ambiguity of $\theta_{23}$ will also affect 
the sensitivity of detecting CP violation. On the other hand, 
we have discussed in our previous communication \cite{MSYIS} 
that the octant degeneracy may be resolved by combining reactor 
measurement of $\theta_{13}$ with the LBL appearance measurement 
in both neutrino and antineutrino channels. 
It would be very interesting to reexamine the possibility in the 
context of this work to clarify to what extent it cures 
the further uncertainty in the sensitivity of detecting CP 
violation mentioned above.

\begin{figure}[htbp]
\includegraphics[width=0.7\textwidth]{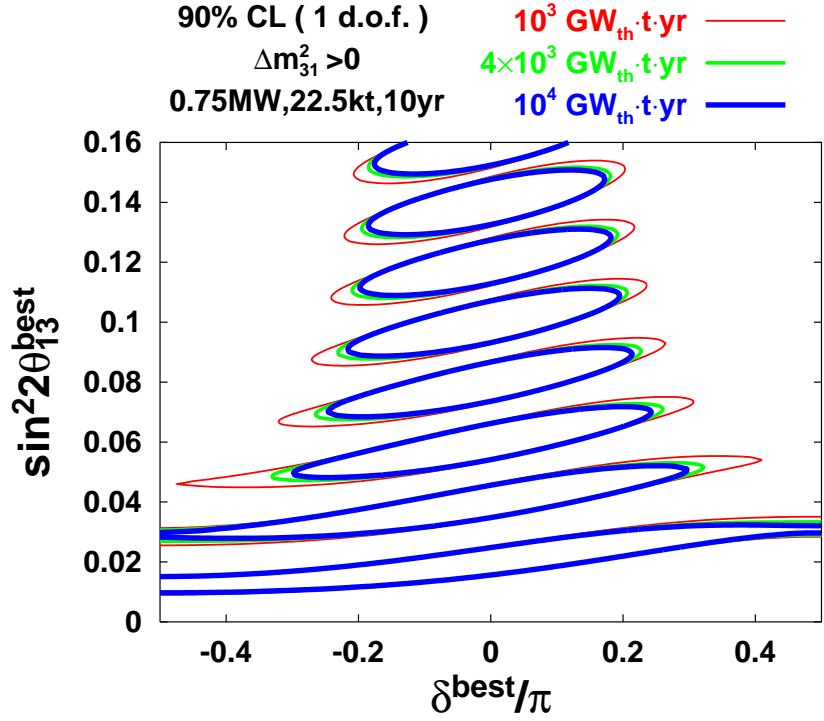}
\caption{
The same as in Fig.~\ref{CPn} but with measurement in 10~years 
running of the JPARC-SK experiment with 0.75MW beam power and 
22.5kt detector (Super-Kamiokande). 
 Although each contour becomes thicker because of a factor of 
$\simeq$~25 lower statistics of the experiment,
the sensitivity to CP violation still exists at 90\%~CL\@.
}
\label{CPSK}
\end{figure}

\vskip 0.3cm
\noindent
(2) We have examined a pessimistic scenario to run the 
JPARC-SK experiment with 0.75MW proton beam power and 
the fiducial volume of 22.5kt, 
while waiting for the construction of Hyper-Kamiokande.
As is shown in Fig.~\ref{CPSK} the sensitivity to CP violation 
becomes worse but still remains for its 10~years running.

\vskip 0.3cm
\noindent
(3) The reactor experiment described in this paper may be 
regarded as the phase II of the currently proposed reactor 
experiments for measuring $\theta_{13}$ 
\cite{krasnoyarsk,kashkari,diablo}, and how to improve the 
systematic errors should be carefully investigated during 
running the phase I experiments.
If it is possible to significantly improve the systematic 
errors over those given in Table~\ref{syserror},  it may be 
possible to extend the CP sensitivity to the region 
$\sin^2{2\theta_{13}} \leq 0.03$.

\vskip 0.3cm
\noindent
(4) From \mbox{Figs.~\ref{CPn}-\ref{CPd}}, it is likely that detection of 
CP violation requires $\sim 10^{3}$ $\GWth\cdot\mbox{ton}\cdot\mbox{year}$ 
measurement by the reactor experiment. 
Now there is a choice between two options: 
stronger power source with smaller detectors, or 
weaker power source with larger detectors. 
If these is no natural or existing holes with enough overburden 
for the detectors the first option might be more advantageous 
because larger detectors require deeper hole to keep the 
signal to noise ratio equal.

\begin{acknowledgments}
We thank Fumihiko Suekane and Osamu Yasuda for many valuable 
discussions and comments.  
We have enjoyed useful conversations and correspondences with 
Takashi Kobayashi, Kenji Kaneyuki, Yoshihisa Obayashi and Masato Shiozawa.
Stephen Parke made useful comments which triggered the revision of 
out treatment of parameter degeneracy due to the $\Delta m^2$ sign.
H.M. is grateful to Theoretical Physics Department of Fermilab 
for hospitality during the Summer Visitor's Program 2003. 
This work was supported by the Grant-in-Aid for Scientific Research
in Priority Areas No. 12047222, Japan Ministry
of Education, Culture, Sports, Science, and Technology.
\end{acknowledgments}


\appendix
\section{Cancellation of errors by near-far detector comparison}

This appendix is meant to be a pedagogical note in which we 
try to clarify the feature of cancellation of systematic errors by 
near-far detector comparison and the relationship between 
over-all and bin-by-bin errors.

The definition of $\Delta\chi^2$ for two detector system is
\begin{eqnarray}
 \Delta\chi_{nf}^2
&\equiv& \min_{\alpha} \Delta\chi_{nf}^2 (\alpha)\nonumber\\
&\equiv& \min_{\alpha} \left[
          \frac{ \left\{ N_f - ( 1 + \alpha ) N_f^\best \right\}^2 }
               { N_f^\best + \sigd^2 (N_f^\best)^2 }
          + \frac{ \left\{ N_n - ( 1 + \alpha ) N_n^\best \right\}^2 }
                { N_n^\best + \sigd^2 (N_n^\best)^2 }
          + \frac{ \alpha^2 }{ \sigD^2 } \right],
\label{chinf}
\end{eqnarray}
where $N_f$ ($N_n$) is the theoretical total number of events
expected to be measured at far (near) detector. 
The quantities with superscript ``best'' are defined as the ones 
calculated with the best-fit values of the ``experimental data'', 
which are to be tested against the CP conserving case.
$\sigD$ and $\sigd$ are correlated and uncorrelated errors
between detectors, respectively. 

We discuss statistical average of an observable $O$ by 
the Gaussian probability distribution function as 
\begin{eqnarray}
< O > \equiv C \int dN_f dN_n d\alpha\, O
                \exp\left( -\frac{1}{\,2\,}\Delta\chi_{nf}^2(\alpha) \right),
\label{integral}
\end{eqnarray}
where $C$ is the normalization constant to make $< 1 >$ unity.
 Note that the integration with respect to $\alpha$
is equivalent to the minimization in (\ref{chinf}).
After the minimization, it takes the following form which is generic to 
the Gaussian distribution, 
\begin{eqnarray}
\Delta\chi_{nf}^2
&=& \left( x, y \right)
     \left(
      \begin{array}{cc}
       <x^2> & <xy> \\
       <yx>  & <y^2>
      \end{array}
     \right)^{-1}
     \left(
      \begin{array}{c}
       x\\
       y
      \end{array}
     \right),
\label{gaussian}
\\[5mm]
x &\equiv& \frac{ N_f - N_f^\best }{N_f^\best}, \ \
y \equiv \frac{ N_n - N_n^\best }{N_n^\best}.
\end{eqnarray}

In order to examine the feature of near-far cancellation of errors,
it is valuable to transform $x$ and $y$ as
\begin{eqnarray}
 X  \equiv x - y = \frac{N_f}{N_f^\best} - \frac{N_n}{N_n^\best}, \ \ \
 Y \equiv x + y = \frac{N_f}{N_f^\best} + \frac{N_n}{N_n^\best} - 2.
\label{trans1}
\end{eqnarray}
Then, $\Delta\chi_{nf}^2$ can be written as in the form (\ref{gaussian})
with 
\begin{eqnarray}
<X^2> &=& \frac{1}{N_f^\best} + \frac{1}{N_n^\best} + 2 \sigd^2, 
\label{X^2} \\
<Y^2> &=& \frac{1}{N_f^\best} + \frac{1}{N_n^\best} + 2 \sigd^2 + 4 \sigD^2,
\label{Y^2} \\
<XY> &=& <YX> = \frac{1}{N_f^\best} - \frac{1}{N_n^\best}.
\end{eqnarray}
It is evident in (\ref{X^2}) that the correlated systematic errors 
cancel by the near-far comparison. 
The systematic error $\sqrt{2}\sigd$ in $<X^2>$ is
referred in \cite{MSYIS,bugey} as the relative normalization error.

We briefly treat the case of two bins with infinite statistics 
to illustrate the importance of uncorrelated errors. In this case 
$X$ subspace of $\Delta\chi^2_{nf}$ can be written as 
\begin{eqnarray}
X (2\sigma_{d}^2)^{-1}X \ \longrightarrow \ 
\left( X_1, X_2 \right)
     \left(
      \begin{array}{cc}
       2\sigdB^2 + 2\sigdb^2 & 2\sigdB^2 \\
       2\sigdB^2  & 2\sigdB^2 + 2\sigdb^2
      \end{array}
     \right)^{-1}
     \left(
      \begin{array}{c}
       X_1\\
       X_2
      \end{array}
     \right).
\end{eqnarray}
It is clear that $\sigdb = 0$ leads to the diverge of 
$\Delta\chi_{nf}^2$ except for the best fit point ($X_i=Y_i=0$), 
which means that the infinite precision can be achieved for the case.
Thus, $\sigdb$ must be treated with great care.

Next we derive the relationship between over-all and bin-by-bin 
errors that was used in the text, (\ref{error_rel}).
For simplicity, we consider the case of one detector with two bins.
Then, $\Delta\chi^2$ for the case is defined as
\begin{eqnarray}
\Delta\chi_{12}^2
&\equiv& \min_{\alpha} \Delta\chi_{12}^2 (\alpha)\nonumber\\
&\equiv& \min_{\alpha} \left[
          \frac{ \left\{ N_1 - ( 1 + \alpha ) N_1^\best \right\}^2 }
               { N_1^\best + \sigb^2 (N_1^\best)^2 }
          + \frac{ \left\{ N_2 - ( 1 + \alpha ) N_2^\best \right\}^2 }
                { N_2^\best + \sigb^2 (N_2^\best)^2 }
          + \frac{ \alpha^2 }{ \sigB^2 } \right],
\end{eqnarray}
where $N_1$ and $N_2$ are the expected numbers of events
within first and second bins, respectively, and 
$\sigB$ ($\sigb$) denotes the correlated (uncorrelated) error
between bins. 

To obtain the error for the total number of events,
we define
\begin{eqnarray}
x_\tot \equiv \sum_i \frac{N_i - N_i^\best}{N_\tot^\best}, \ \
N_\tot^\best \equiv \sum_i N_i^\best.
\end{eqnarray}
Then, we obtain 
\begin{eqnarray}
< x_\tot^2 > &=& C' \int dN_1 dN_2 d\alpha\, x_\tot^2 
            \exp\left( -\frac{1}{\,2\,}\Delta\chi_{12}^2(\alpha) \right) 
\nonumber \\
&=& 
\frac{1}{N_\tot^\best}
         + \sigB^2 + \sigb^2 \frac{\sum_i (N_i^\best)^2}{(N_\tot^\best)^2}.
\end{eqnarray}
One can show that the same treatment goes though for arbitrary 
number of bins. The coefficient of $\sigb^2$ is almost 1/9
in our analysis (14 bins).




\newpage

\vglue 1.cm

\newpage

\vglue 1.cm

\newpage

\vglue 1.cm

\newpage

\vglue 1.cm

\end{document}